\documentclass[10pt]{article}

\usepackage[leqno]{amsmath}
\usepackage{amssymb}
\usepackage{graphicx}
\usepackage{indentfirst,csquotes}
\usepackage{xcolor,paralist,fancyhdr,etoolbox}
\usepackage{pgfplots}
\usepackage{xurl}
\usepackage{titlesec}
\usepackage[breaklinks=true]{hyperref}
\usepackage[font=small,labelfont=bf]{caption}

\pgfplotsset{compat=1.18}
\hypersetup{
  colorlinks=true,
  linkcolor=black,
  filecolor=black,
  urlcolor=black
}

\titleformat{\section}
  {\normalfont\Large\bfseries}
  {\thesection.}{0.5em}{}

\titleformat{\subsection}
  {\normalfont\large\bfseries}
  {\thesubsection.}{0.5em}{}

\titleformat{\subsubsection}[runin]
  {\normalfont\normalsize\bfseries}
  {\thesubsubsection.}{0.5em}{}

\titlespacing*{\section}{0pt}{2.5ex plus 1ex minus .2ex}{1.5ex}
\titlespacing*{\subsection}{0pt}{2ex plus 0.8ex minus .2ex}{1ex}
\titlespacing*{\subsubsection}{0pt}{1ex}{0.5em}

\makeatletter
\renewcommand{\maketitle}{
  \begin{center}
    {\LARGE\bfseries \@title \par}
    \vspace{1.1em}
    {\normalsize \@author \par}
    \vspace{0.6em}
    {\small \@date \par}
  \end{center}
  \vspace{1.2em}
}
\makeatother

\begin{document}

\title{The Cost of Quantum Resistance: A Hash-Based Commit--Reveal Alternative for Minimizing Blockchain Infrastructure Overhead}

\author{
Keir Finlow-Bates\textsuperscript{1} \quad
Markus Jakobsson\textsuperscript{1} \quad
Hossein Siadati\textsuperscript{1}\\[0.3em]
\textsuperscript{1}Artema LABS\\[0.3em]
\texttt{\{keir,markus,hossein\}@artemalabs.com}
}

\date{\today}

\maketitle

\begin{abstract}
The transition to post-quantum cryptography in blockchain systems such as Bitcoin and Ethereum is often framed as a purely cryptographic problem. In practice, it also presents significant economic and infrastructural challenges: in globally replicated networks, increases in transaction size and verification cost are multiplied across all participating nodes.

Existing post-quantum signature schemes, including lattice-based constructions such as CRYSTALS-Dilithium and stateless hash-based schemes such as SPHINCS+, introduce substantial increases in signature size. At blockchain scale, these increases translate into higher storage, bandwidth, and validation requirements, potentially requiring multiple generations of hardware improvement to become operationally routine. Historical experience suggests that even moderate increases in data footprint can be contentious, as illustrated by the Bitcoin block size debates (2015--2017).

We propose a hash-based commit--reveal construction that replaces a single signature-bearing transaction with two lightweight transactions, each containing a fixed-size (32-byte) hash output derived from well-established primitives such as SHA-256, BLAKE, or Keccak. This approach achieves post-quantum security under standard hash assumptions while increasing the effective transaction footprint by only approximately 1.5$\times$ to 2$\times$ per authorization event.

These results indicate that practical post-quantum migration may benefit from rethinking transaction semantics rather than directly adopting larger signature schemes, and that viable designs for decentralized systems must account for system-wide cost amplification.
\end{abstract}

\section{Introduction}

The security of contemporary blockchain systems such as Bitcoin~\cite{nakamoto2008} and Ethereum~\cite{buterin2014} is based on elliptic curve cryptography, widely believed to be vulnerable to sufficiently powerful quantum adversaries due to Shor's algorithm~\cite{shor1997}. This has led to growing interest in post-quantum signature schemes as potential replacements for existing transaction authorization mechanisms~\cite{fernandez2020pqblockchain}. However, most discussions of post-quantum migration treat the problem as primarily cryptographic: identifying schemes that resist quantum attacks while maintaining acceptable security parameters.

We argue that this framing is incomplete: in blockchain systems, the dominant constraint is not signature security in isolation, but the global cost of replicating that security across the entire network. Every transaction is stored, transmitted, and verified by a large number of nodes. As a result, increases in signature size or verification complexity are not local costs but system-wide multipliers affecting storage, bandwidth, and computational load across the entire network. Consequently, the feasibility of post-quantum migration depends not only on security properties, but also on the economic and infrastructural impact of candidate schemes~\cite{schemitt2025pqcblockchain}.

Existing post-quantum signature schemes, including lattice-based constructions such as CRYSTALS-Dilithium~\cite{dilithium} and stateless hash-based schemes such as SPHINCS+~\cite{sphincsplus}, provide strong security guarantees under quantum adversaries and are part of the broader NIST standardization effort~\cite{nistpqc}. However, they introduce substantially larger signatures and sometimes higher computational overhead than current elliptic curve schemes. When deployed in globally replicated systems, these increases translate into significant growth in blockchain size and validation costs. Such growth risks exceeding the pace of hardware improvement, requiring multiple generations of storage and processing upgrades to be deployed within a compressed timeframe, rather than over the decade-long horizon typically associated with Moore’s law, before becoming operationally routine.

Historical experience underscores the sensitivity of blockchain systems to
changes in data footprint. The Bitcoin block size debates between 2015 and
2017, culminating in the activation of SegWit and the Bitcoin Cash hard fork,
demonstrated that even modest increases in block capacity can lead to prolonged
contention, ecosystem fragmentation, and competing protocol visions
\cite{bitcoin_scaling_wikipedia,segwit_wikipedia,bitcoin_cash_wikipedia}.
These events highlight that scalability is not merely a technical parameter,
but a socio-economic constraint that influences adoption and consensus.

In this paper, we argue that minimizing total network cost is a first-order requirement for post-quantum blockchain design. Rather than embedding increasingly large signatures into individual transactions, we propose a fundamentally different approach based on a hash-based commit--reveal paradigm. Our construction replaces a single signature-bearing transaction with two lightweight transactions, each containing a fixed-size (32-byte) hash-chain output derived from well-established cryptographic hash functions such as SHA-256, BLAKE, or Keccak.

This approach achieves post-quantum security under standard hash function assumptions while increasing the effective transaction footprint by only approximately 1.5$\times$ to 2$\times$ per authorization event. By relying exclusively on mature and extensively analyzed primitives, the scheme avoids the need for newer algebraic hardness assumptions and reduces implementation complexity and risk.

The key insight is that post-quantum security for blockchains does not necessarily require larger signatures, but rather different transaction semantics. By shifting complexity from signature size to transaction structure, the proposed commit--reveal construction bounds system-wide cost increases and remains compatible with the incremental evolution of hardware and network infrastructure.

The remainder of this paper is structured as follows. In Section~2, we review relevant background on blockchain transaction models, system-wide cost amplification, and post-quantum signature schemes. Section~3 examines the economic and infrastructural implications of direct post-quantum migration, including storage growth, hardware requirements, and network-wide cost estimates. Section~4 then presents the proposed commit--reveal authorization scheme in detail, including its construction, security intuition, and implications for practical deployment, and Section~5 concludes.

\section{Background}

\subsection{Blockchain Cost Model}

Blockchain systems such as Bitcoin and Ethereum operate as globally replicated ledgers in which every transaction is broadcast, validated, and stored by a distributed network of nodes. Unlike centralized systems, where data is stored once and accessed on demand, blockchains impose a replication requirement: each participating node maintains a local copy of the ledger sufficient to independently verify its correctness.

As a result, the cost of processing a transaction is not confined to the party submitting it, but is distributed across the entire network. Each additional byte included in a transaction must be transmitted over the peer-to-peer network, validated by nodes, and persisted to disk for long-term storage. This leads to a fundamental cost amplification effect in which local increases in transaction size translate into global increases in resource consumption.

A useful abstraction is to model the system-wide cost of a transaction as scaling proportionally with both its size and the number of nodes maintaining the ledger. That is,
\[
\text{Total Cost} \propto \text{Transaction Size} \times \text{Number of Nodes}.
\]
While this expression omits constant factors and differences in node roles, it captures the key intuition that blockchain systems convert per-transaction data overhead into network-wide resource requirements.

This cost model has direct implications for cryptographic design. In conventional systems, increasing the size of a digital signature affects only local storage or communication overhead. In contrast, in a blockchain setting, larger signatures increase the global burden on storage, bandwidth, and validation. Consequently, data size becomes a first-order design constraint rather than a secondary implementation detail.

Importantly, economic pricing mechanisms, such as transaction fees or gas costs, do not eliminate this underlying cost. They merely determine how the burden is distributed among participants. Regardless of pricing, the physical data must still be transmitted and stored by the network. This distinction between accounting cost and physical cost is central to understanding the challenges of deploying post-quantum cryptography in blockchain environments.

\subsection{Bitcoin Transactions and Size}

Standard Bitcoin transactions represent the transfer of value by consuming previously unspent transaction outputs (UTXOs) and creating new ones. Each transaction consists of a set of inputs, a set of outputs, and associated metadata. Inputs reference prior outputs and include authorization data, while outputs specify recipient addresses and amounts.

A typical Bitcoin transaction comprises the following components:
\begin{itemize}
    \item Version and locktime fields
    \item One or more inputs, each containing:
    \begin{itemize}
        \item A reference to a previous transaction output
        \item A script (scriptSig or witness data) containing the authorization information
    \end{itemize}
    \item One or more outputs, each containing:
    \begin{itemize}
        \item A value in satoshis
        \item A locking script (scriptPubKey) defining spending conditions
    \end{itemize}
\end{itemize}

The size of a Bitcoin transaction depends on the number of inputs and outputs, but typical transactions fall in the range of approximately 200 to 300 bytes for legacy (pre-SegWit) formats. A commonly cited example is a transaction with one input and two outputs. For a standard legacy transaction, this can be estimated as follows:
\begin{itemize}
    \item Base fields (version, locktime, counts): $\sim$10 bytes
    \item Input: $\sim$148 bytes
    \item Outputs (2 $\times$ 34 bytes): $\sim$68 bytes
\end{itemize}
giving a total size of approximately 226 bytes\cite{bitcoin_wiki_tx}.

A significant portion of this size is attributable to the digital signature and associated public key used to authorize each input. For transactions using the Elliptic Curve Digital Signature Algorithm (ECDSA) over the secp256k1 curve, each input typically includes a signature of approximately 64--72 bytes and a public key of approximately 33 bytes, along with additional script overhead. As a result, authorization data constitutes a substantial fraction of the total transaction size.

Because Bitcoin uses the UTXO model, transactions with multiple inputs scale linearly in size: each additional input requires its own signature and associated data. Consequently, transactions that consolidate multiple UTXOs can be significantly larger, further amplifying the contribution of signature material to overall transaction size.

From the perspective of the cost model introduced in Section~2.1, this structure implies that signature data is a dominant factor in determining the global resource cost of Bitcoin transactions. Any increase in signature size, such as that introduced by post-quantum signature schemes, therefore has a direct and multiplicative impact.

\subsection{SegWit and Witness Data}

Segregated Witness (SegWit), activated in Bitcoin in 2017, introduced a
modification to the transaction format that separates authorization data from
the transaction body. In SegWit transactions, signatures and related unlocking
data are moved into a distinct structure known as the witness, while the
traditional transaction identifier remains computed over the serialization
without witness data \cite{bip141Segwit,bitcoinCoreSegwitWalletGuide}.

This change was motivated in part by the need to increase effective block
capacity without a direct hard-fork increase in the nominal 1 MB block-size
limit. BIP141 replaced the old block-size rule with a block-weight rule: base
transaction data is counted at full weight, while witness data is effectively
discounted. More precisely, block weight is defined from the base size and total
serialized size, and blocks are limited to \(4{,}000{,}000\) weight units
\cite{bip141Segwit}. The same accounting gives rise to virtual transaction size
or vbytes, where witness bytes contribute one quarter as much as non-witness
bytes for fee and block-capacity accounting
\cite{bip141Segwit,bitcoinOptechTransactionSizeCalculator}.

However, this distinction between virtual size and actual size is an accounting
mechanism rather than a reduction in physical data. BIP141 defines the total
serialized size of a block as including both base data and witness data, and
SegWit-capable nodes relay transactions using the serialization that includes
the witness fields \cite{bip141Segwit,bip144SegwitPeerServices,
bitcoinCoreSegwitWalletGuide}. Thus, although witness data is discounted for
the purposes of block weight and fee calculation, it is still transmitted,
validated, and stored by fully validating nodes.

This leads to an important observation: reducing the \emph{priced} cost of
signatures does not reduce their \emph{physical} cost. Even when signature data
is moved into a discounted region of the block, it continues to contribute to
the global storage and bandwidth requirements of the network. In other words,
SegWit changes how signature data is accounted for, but not the fact that it
must be replicated and persisted by full nodes.

This distinction has direct implications for post-quantum cryptography.
Although it is in principle possible to assign lower virtual costs to larger
post-quantum signatures, such accounting adjustments do not eliminate the
underlying resource burden.

\subsection{Ethereum Transaction Model (Externally Owned Accounts)}

Ethereum, in contrast, employs an account-based model in which transactions directly modify account state rather than consuming and creating unspent outputs as in Bitcoin. Two types of accounts exist: externally owned accounts (EOAs), which are controlled by private keys, and smart contract accounts, which are controlled by code. This work focuses on transactions originating from EOAs, as these represent the most common transaction type and the primary use of digital signatures in Ethereum.

A typical Ethereum transaction includes the following components:
\begin{itemize}
    \item Nonce (transaction sequence number)
    \item Gas price or fee parameters
    \item Gas limit
    \item Recipient address
    \item Value (amount of Ether transferred)
    \item Data payload (optional)
    \item Signature fields $(v, r, s)$
\end{itemize}

The signature fields $(r, s)$ correspond to a standard ECDSA signature over the secp256k1 curve, while $v$ encodes recovery and chain information. The combined size of the signature is typically 65 bytes, which forms a significant portion of the serialized transaction. Overall, a simple Ether transfer transaction typically occupies on the order of 100 to 200 bytes, depending on encoding and parameter sizes. This estimate follows from the fixed-size signature fields (65 bytes) and the 
RLP-encoded transaction fields specified in the Ethereum protocol~\cite{wood2014ethereum}.

Unlike Bitcoin, Ethereum does not explicitly price transaction data by byte. Instead, transactions incur a base gas cost (currently 21{,}000 units for a simple transfer), which implicitly includes the cost of signature verification. This abstraction decouples transaction fees from the exact size of the signature, allowing signatures to be validated without a direct per-byte pricing model.

However, as with Bitcoin and SegWit, this distinction between execution cost and data size does not eliminate the underlying storage requirement. Transaction data, including signatures, is included in blocks and must be stored by at least some of the nodes. While the gas model determines the immediate cost to the sender, it does not change the fact that larger signatures increase the total amount of data that must be propagated and persisted across the network.

Furthermore, Ethereum supports additional transaction patterns in which signatures are supplied as part of calldata to smart contracts rather than as part of the transaction envelope itself. In these cases, signature data is explicitly charged per byte, further emphasizing the sensitivity of the system to increases in signature size. Such contract-level mechanisms are outside the scope of this work.

The key observation is that, as in Bitcoin, signature data constitutes a non-trivial component of transaction size and contributes directly to the global resource cost of the system. Consequently, replacing ECDSA signatures with substantially larger post-quantum alternatives would increase both the immediate and long-term resource burden of Ethereum, regardless of how those costs are accounted for in the fee model.

\subsection{Node Types and Storage Persistence}

Blockchain networks support different types of nodes with varying storage and verification responsibilities. In Ethereum, nodes are commonly categorized as light nodes, full nodes, and archive nodes, each reflecting a different trade-off between resource usage and independence.

Light nodes store only a minimal subset of blockchain data, typically retaining recent block headers and relying on other nodes to provide additional information on demand. As a result, they do not maintain a complete history of transactions or state and are able to operate with significantly reduced storage requirements.

Full nodes, by contrast, store all transaction data necessary to independently verify the correctness of the blockchain. This includes the full history of transactions and their associated data, including digital signatures. Archive nodes extend this further by preserving every historical state transition, allowing reconstruction of the entire state of the system at any point in time. These nodes represent the highest storage burden in the network.

The security of the network depends critically on the existence of full nodes and archive nodes, which provide independent verification and historical integrity. While light nodes can discard or prune data, full and archive nodes must retain transaction data, including signatures, for extended periods or indefinitely. Consequently, the cumulative storage cost of the blockchain is borne by this subset of nodes.

This distinction has important implications for the evaluation of post quantum signature schemes. Even if certain participants in the network can avoid storing large signatures through pruning or reliance on external data, the system as a whole must still accommodate their permanent storage. Larger signatures therefore impose a persistent and growing burden on full and archive nodes, increasing hardware requirements, and potentially raising the barrier to entry for independent participation.

From the perspective of decentralization, this effect is particularly significant. As storage and bandwidth requirements increase, fewer participants may be willing or able to operate full nodes, leading to greater reliance on centralized infrastructure. Thus, increases in signature size do not merely affect performance, but can also influence the long-term structure and resilience of the network.

\subsection{Post-Quantum Signature Schemes}

The vulnerability of elliptic curve cryptography to quantum attacks has motivated the development of post-quantum digital signature schemes designed to remain secure against adversaries equipped with quantum computers. These schemes rely on alternative hardness assumptions, including lattice problems, hash-based constructions, and multivariate polynomial systems. Among the most prominent candidates are CRYSTALS-Dilithium, a lattice-based signature scheme, and SPHINCS+, a stateless hash-based scheme, both of which have been selected as part of the NIST post-quantum standardization process.

While these schemes provide strong security guarantees, they introduce substantially larger key and signature sizes compared to classical schemes such as ECDSA. Typical parameter sizes illustrate the magnitude of this difference. An ECDSA signature over the secp256k1 curve is approximately 65 bytes in size. In contrast, Dilithium signatures range from approximately 2~KB to 4.5~KB depending on the security level, while SPHINCS+ signatures are typically on the order of tens of kilobytes. Earlier hash-based constructions such as Lamport and Winternitz one-time signatures also produce signatures in at least the tens of kilobytes range.

\begin{table}[h]
\centering
\begin{tabular}{l c c}
\hline
\textbf{Scheme} & \textbf{Public Key Size} & \textbf{Signature Size} \\
\hline
ECDSA (secp256k1) & $\sim$64 bytes & $\sim$65 bytes \\
Dilithium (level 2--5) & $\sim$1--2.5 KB & $\sim$2--4.5 KB \\
SPHINCS+ & $\sim$32 bytes & $\sim$10--30 KB \\
Lamport (hash-based) & $\sim$32 KB & $\sim$16 KB \\
\hline
\end{tabular}
\caption{Representative public key and signature sizes for classical and post-quantum schemes.}
\end{table}

These differences are not marginal but represent increases of one to two orders of magnitude in signature size. It is important to note that these schemes are not impractical in general; rather, they are misaligned with the constraints of blockchain environments. The primary challenge is not computational feasibility, but data footprint.

Consequently, while post-quantum signature schemes such as Dilithium and SPHINCS+ are cryptographically sound, their direct adoption in blockchain systems raises significant scalability concerns. This tension between security and system-wide cost motivates the exploration of alternative approaches that achieve quantum resistance without incurring substantial increases in on-chain data size.

\subsection{System-Wide Cost Amplification}

\begin{figure}[h]
\centering
\begin{tikzpicture}
\begin{axis}[
    ybar,
    bar width=14pt,
    width=11cm,
    height=6.5cm,
    symbolic x coords={ECDSA, Dilithium, SPHINCS+},
    xtick=data,
    ylabel={Transaction Size (bytes)},
    ymin=0,
    ymax=30000,
    scaled y ticks=false,
    nodes near coords,
    nodes near coords align={vertical},
]
\addplot coordinates {
    (ECDSA,200)
    (Dilithium,3000)
    (SPHINCS+,20000)
};
\end{axis}
\end{tikzpicture}
\caption{Approximate transaction size under different signature schemes. Post-quantum signatures increase transaction size by one to two orders of magnitude.}
\label{fig:tx_size}
\end{figure}
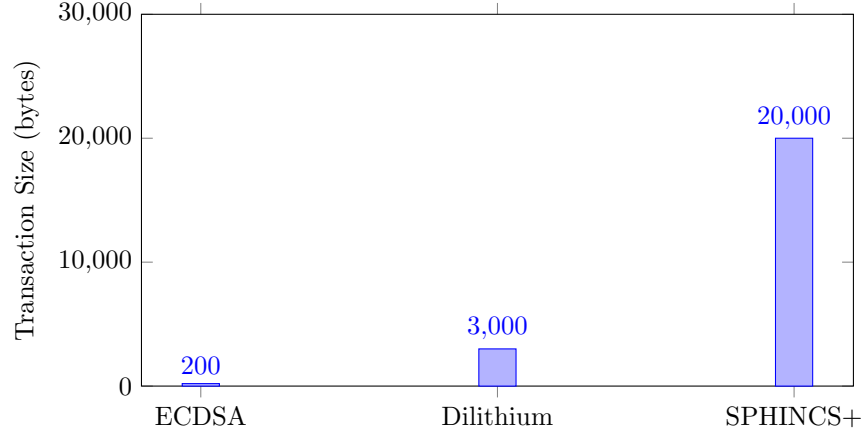

The preceding sections highlight a consistent theme across both Bitcoin and Ethereum: blockchain systems transform local data overhead into global resource consumption. As shown in Section~2.6, post-quantum signature schemes introduce signatures that are one to two orders of magnitude larger than those used in current systems. Figure~\ref{fig:tx_size} illustrates the approximate effect on transaction size for representative signature schemes.

As discussed in Section~2.3, techniques such as SegWit can adjust the accounting cost of certain types of data, but do not eliminate the need to physically store and propagate that data. Similarly, Ethereum's gas model can abstract away per-byte pricing for certain operations, but does not change the underlying storage burden imposed on full and archive nodes.

The consequence is a fundamental tension between post-quantum security and blockchain scalability. Direct substitution of existing signature schemes with larger post-quantum alternatives increases the global cost of operating the network, potentially raising hardware requirements and reducing accessibility for independent node operators. This, in turn, will impact decentralization by increasing reliance on well-provisioned infrastructure.

From this perspective, the challenge of post-quantum migration in blockchain systems is not solely one of cryptographic design, but of system-wide cost control. Any viable solution must account for the amplification effect inherent in globally replicated ledgers and seek to minimize the additional data burden introduced by quantum-resistant mechanisms.

\subsection{Storage Growth and Hardware Implications}

Current blockchain storage requirements already impose substantial hardware demands on node operators. As of 2026, the Bitcoin blockchain exceeds approximately 700~GB in size~\cite{bitcoin_size}, while Ethereum full nodes require on the order of 1--1.5~TB of storage~\cite{eth_full_node_size}. Archive nodes, which retain the full historical state, may require more than 10~TB depending on client implementation and configuration~\cite{eth_archive_node_size}. These values are indicative and may vary with pruning strategies, synchronization modes, and software clients, but they establish the relevant order of magnitude.

Post-quantum signature schemes would significantly increase these requirements. For example, replacing ECDSA signatures (approximately 65 bytes) with SPHINCS+ signatures (typically 10--30~KB) introduces an increase of two orders of magnitude in signature size. Given that signatures constitute a substantial portion of transaction data, this increase propagates directly into overall transaction size.

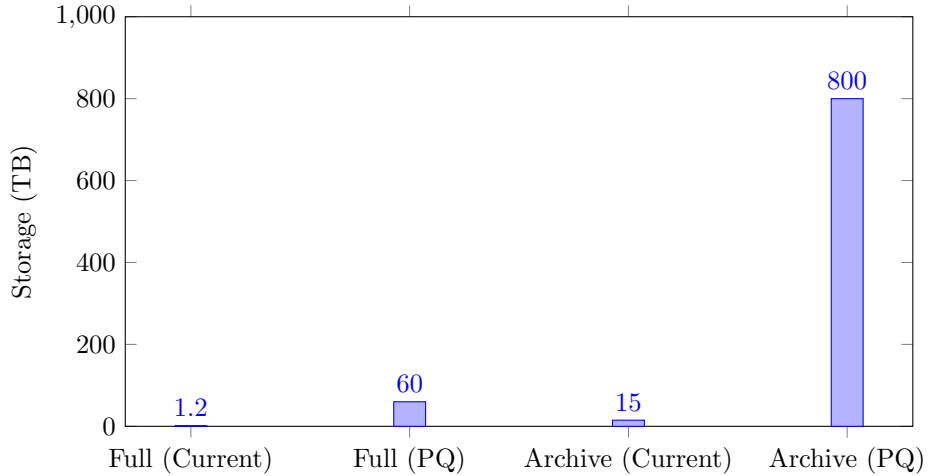
\begin{figure}[h]
\centering
\begin{tikzpicture}
\begin{axis}[
    ybar,
    bar width=12pt,
    width=12cm,
    height=7cm,
    symbolic x coords={Full (Current), Full (PQ), Archive (Current), Archive (PQ)},
    xtick=data,
    ylabel={Storage (TB)},
    ymin=0,
    ymax=1000,
    nodes near coords,
    nodes near coords align={vertical},
]
\addplot coordinates {
    (Full (Current),1.2)
    (Full (PQ),60)
    (Archive (Current),15)
    (Archive (PQ),800)
};
\end{axis}
\end{tikzpicture}
\caption{Projected storage requirements for Ethereum nodes under post-quantum signature schemes. Archive nodes approach the petabyte scale (1000~TB).}
\label{fig:eth_storage}
\end{figure}

Under conservative assumptions, this would increase Bitcoin transaction sizes by approximately one to two orders of magnitude, implying a corresponding increase in total blockchain size from hundreds of gigabytes to tens of terabytes. For Ethereum, where transaction sizes are smaller relative to signature payloads, the proportional impact is at least as large, potentially increasing full node storage requirements from approximately 1~TB to tens of terabytes, and archive node requirements into the hundreds of terabytes. Figure~\ref{fig:eth_storage} illustrates the scale of this projected increase for Ethereum nodes.

Such increases represent a qualitative shift in infrastructure requirements rather than a marginal extension of current trends. Assuming typical hardware improvement rates, such as storage capacity doubling approximately every two years, accommodating a 50$\times$--100$\times$ increase in data volume would require an almost immediate advancement across multiple hardware generations corresponding to a time horizon on the order of a decade or more. This introduces a temporal mismatch between cryptographic transition and infrastructure capability.

\section{Economic and Infrastructure Implications}

\subsection{Hardware Requirements and Scaling Limits}

Current blockchain node requirements already approach terabyte-scale storage, as discussed in Section~2.8. Direct adoption of post-quantum signature schemes would increase transaction sizes by one to two orders of magnitude, leading to corresponding increases in total chain size. Assuming typical hardware improvement trends, such as storage capacity doubling approximately every two years at the same price point, accommodating a 50$\times$--100$\times$ increase in data volume would require almost immediate advancement across multiple hardware generations, corresponding to a time horizon on the order of 10--15 years.

\subsection{Network-Wide Upgrade Costs}

To estimate the infrastructure implications of direct post-quantum migration, we consider the effect of replacing ECDSA with a large stateless hash-based signature scheme such as SPHINCS+ under a simplified, order-of-magnitude model. The purpose of this analysis is not to produce precise forecasts, but to characterize the scale of the resulting storage burden under plausible assumptions.

\paragraph{Node Counts.}
For Bitcoin, public network measurements distinguish between publicly reachable
nodes and additional nodes inferred from peer-address gossip. Bitnodes estimates
the global Bitcoin peer-to-peer network, including both reachable and unreachable
nodes, at approximately \(72{,}000\) nodes, while contemporaneous reachable-node
crawls observe roughly \(13{,}000\)--\(20{,}000\) reachable full nodes
\cite{bitnodesGlobalBitcoinNodes}. Academic measurement work similarly emphasizes
that reachable nodes represent only the visible portion of the Bitcoin network,
and that unreachable nodes may still participate in validation, storage, and
message propagation \cite{li2023bns}. Since replicated storage costs are borne
by validating nodes rather than only by publicly reachable nodes, we use
\(\sim 72{,}000\) as an estimate of total participating Bitcoin
nodes.

For Ethereum, public network monitors report node counts on the order of
\(10^4\): Etherscan's node tracker reports approximately \(13{,}000\)
detected Ethereum nodes, while Ethernodes reports approximately \(7{,}000\)
execution-layer clients and \(8{,}000\) consensus-layer clients
\cite{etherscanEthereumNodeTracker,ethernodesEthereumClients}.  Ethereum
archive nodes represent a more storage-intensive configuration: an archive node
stores historical states, whereas a full node verifies and follows the latest
state and prunes older intermediate state data
\cite{ethereumArchiveNodes}.  Since public monitors do not provide a
standardized global split between full and archive nodes, we use the following
order-of-magnitude modelling assumptions:
\begin{itemize}
    \item \(\approx 12{,}000\) Ethereum full nodes,
    \item \(\approx 1{,}000\) Ethereum archive nodes.
\end{itemize}

Several smaller UTXO-based networks also contribute to the replicated storage
footprint. Publicly accessible node crawlers indicate node counts on the order
of:
\begin{itemize}
    \item Litecoin: \(\sim 10^3\),
    \item Dogecoin: \(\sim 5 \times 10^2\),
    \item Bitcoin Cash: \(\sim 5 \times 10^2\),
    \item Bitcoin SV: \(\sim 10^2\)--\(10^3\), treated as an
    order-of-magnitude assumption due to the lack of a standardized public
    global node-count monitor.
\end{itemize}
These values are based on public crawler snapshots for Litecoin, Dogecoin, and
Bitcoin Cash \cite{blockchairLitecoinNodes,blockchairDogecoinNodes,coinDanceBCHNodes}.

In addition, the Ethereum ecosystem includes Layer-2 systems such as Arbitrum,
Optimism, Base, zkSync, and Starknet. These systems are tracked as major
Ethereum scaling protocols by public Layer-2 analytics services, but reliable
global node counts for their supporting infrastructure are not publicly
standardized \cite{l2beat}. We therefore treat Layer-2 contributions as a
potentially significant but unquantified extension to the estimates below.

\paragraph{Baseline Chain Growth.}
As a coarse baseline, we assume approximate annual chain growth of:
\begin{itemize}
    \item Bitcoin: $\sim 50$--$70$ GB/year,
    \item Ethereum full nodes: $\sim 150$--$250$ GB/year.
\end{itemize}

\paragraph{Signature-Induced Growth.}
For an order-of-magnitude upper-bound estimate, we approximate the effect of
large post-quantum signatures by applying a \(50\times\) multiplier to baseline
chain growth. This intentionally over-aggregates non-signature transaction data
and should therefore be interpreted as a stress-case rather than a precise
forecast.

\paragraph{Aggregate Replicated Storage.}
Multiplying these per-node increases by the corresponding node populations gives
an approximate total additional replicated storage burden. Table~\ref{tab:pq-storage-burden}
summarizes the resulting order-of-magnitude estimates for the main networks
considered. This subtotal excludes Layer-2 systems, whose inclusion would increase the
overall replicated storage requirement further.

\begin{table}[h]
\centering
\small
\begin{tabular}{p{2.9cm} p{1.9cm} p{2.4cm} p{2.8cm}}
\hline
\raggedright\textbf{Network} &
\centering\textbf{\shortstack{10-year\\growth}} &
\centering\textbf{\shortstack{Additional data\\per node}} &
\centering\textbf{\shortstack{Additional total\\storage}} \tabularnewline
\hline
\raggedright Bitcoin &
\centering $\sim 0.6$ TB &
\centering $\sim 30$ TB &
\centering $\sim 2.2$ EB \tabularnewline

\raggedright Ethereum (full) &
\centering $\sim 2$ TB &
\centering $\sim 100$ TB &
\centering $\sim 1.2$ EB \tabularnewline

\raggedright Ethereum (archive) &
\centering $\sim 20$ TB &
\centering $\sim 1$ PB &
\centering $\sim 1$ EB \tabularnewline

\raggedright Other UTXO chains &
\centering --- &
\centering --- &
\centering $\sim 0.06$--$0.09$ EB \tabularnewline
\hline
\raggedright\textbf{Subtotal} &
\centering --- &
\centering --- &
\centering $\sim 4.5$ EB \tabularnewline
\hline
\end{tabular}
\caption{Approximate ten-year storage growth and aggregate replicated storage burden under a simplified post-quantum signature migration model.}
\label{tab:pq-storage-burden}
\end{table}

\paragraph{Interpretation.}
Under these assumptions, direct adoption of large post quantum signatures would add multiple exabytes of replicated storage across major blockchain ecosystems over a decade. The point is not the precise value of any individual estimate, but the magnitude of the shift: the resulting burden is qualitatively different from ordinary incremental chain growth and instead approaches internet-scale infrastructure expansion.

These estimates are intentionally simplified. They rely on approximate node counts, coarse growth models, and an assumed signature-size multiplier, and they implicitly assume that networks preserve existing throughput rather than accommodating larger signatures by reducing activity or increasing fees.

\subsection{Infrastructure Cost Model}

The storage estimates derived above quantify the additional data volume implied by large post-quantum signatures. Translating this volume into real-world impact requires considering how storage is provisioned and operated in practice.

\paragraph{Baseline Media Cost.}
At enterprise SSD pricing of approximately \$300/TB, the additional $\sim 4.5$ exabytes of replicated storage correspond to a lower-bound raw media cost of:
\[
\approx \$1.35\ \text{billion}.
\]
This figure reflects only the cost of storage media under idealized assumptions and does not capture the full infrastructure burden.

\paragraph{Provisioning Overheads.}
In deployed systems, logical storage requirements must be translated into resilient, maintainable infrastructure. This introduces several multiplicative cost factors:
\begin{itemize}
    \item \textbf{Usable capacity overhead ($\sim 1.3\times$):} headroom, filesystem overhead, and indexing,
    \item \textbf{Redundancy ($\sim 2\times$):} replication and redundancy, for example, RAID,
    \item \textbf{Storage system overhead ($\sim 2$--$2.5\times$):} servers, enclosures, controllers, memory, and networking,
    \item \textbf{Deployment and integration ($\sim 1.5\times$):} installation, provisioning, and data migration,
    \item \textbf{Lifecycle costs ($\sim 1.5\times$):} hardware replacement and failure over a multi-year horizon.
\end{itemize}

Combining these factors yields an overall multiplier in the range of approximately $10\times$ to $20\times$ over raw media cost. Applying this to the baseline estimate gives:

\[
\text{Total cost} \approx (10\text{--}20) \times \$1.35\ \text{B}
\approx \$14\text{--}27\ \text{B}.
\]

\paragraph{Interpretation.}
While such aggregate costs are modest relative to global technology or
financial-sector investment, they should be understood differently from the
capital and operating costs of centralized financial infrastructure. For example,
SWIFT serves more than \(11{,}500\) institutions across over 200 countries and
territories, supported by a centralized organizational and revenue model
\cite{swift_institutions_2024,swift_revenue_2024}. Public blockchain storage
costs, by contrast, are distributed across a diffuse set of node operators, most
of whom receive no direct economic compensation for storing and validating
blockchain data.

From this perspective, the relevant constraint is not aggregate capital availability, but the willingness of marginal participants to incur these costs. Even at the level of individual operators, accommodating tens of terabytes of additional data may require expenditures on the order of thousands to tens of thousands of dollars, which is non-trivial in the absence of direct financial return.

\paragraph{Implications for Participation and Decentralization.}
An important feature of existing blockchain systems is that full node operation is often voluntary and weakly incentivized. As infrastructure requirements increase, participation may become concentrated among well-resourced entities such as exchanges, custodians, and infrastructure providers.

This introduces a structural tension: while the network may remain operational, the distribution of validation becomes less decentralized. In this context, the cost increases associated with large post-quantum signatures are not merely an economic consideration, but a potential driver of structural centralization.

\section{Commit--Reveal Authorization Scheme}

\subsection{Design Rationale}

As established in Section~2, directly embedding post-quantum signatures into blockchain transactions leads to substantial increases in transaction size and system-wide cost. This motivates an alternative approach in which authorization is achieved without transmitting large signature payloads.

We propose a commit--reveal authorization scheme for individual transaction
requests, which are responsible for the vast majority of signature-related data
storage requirements. The scheme replaces traditional digital signatures with a
two-step protocol based on preimages of one-way functions, following the
hash-based secure-transaction construction described in
\cite{FinlowBatesJakobsson2025SecureTransactions}. The scheme relies only on
standard cryptographic hash functions and avoids the need for large signature
objects or complex algebraic constructions.

At a high level, authorization is expressed as the successful revelation of a secret value whose hash was previously committed on-chain. This transforms the problem of digital signing into one of demonstrating knowledge of a preimage, while maintaining compatibility with existing blockchain transaction models.

\subsection{Basic Construction}

Let $F:\{0,1\}^* \rightarrow \{0,1\}^n$ be a one-way function, instantiated in practice by a cryptographic hash function such as SHA-256, BLAKE, or Keccak.

A user generates a random secret preimage $x \in \{0,1\}^n$ and computes:
\[
y = F(x)
\]

The value $y$ serves as a public authorization identifier (analogous to a public key or address), while $x$ remains secret.

To authorize an action \(m\), such as a transfer of assets or a smart contract
invocation, the user performs the following two-step protocol. This framing is
compatible with account-based and contract-oriented blockchain execution models,
in which externally submitted transactions trigger state transitions and smart
contract calls \cite{FinlowBates2026SmartContractInnovation}.

\subsubsection{Commit Transaction}
The user submits a transaction containing a commitment:
\[
C = \big( F(x), \; F(x \parallel m) \big)
\]
where $\parallel$ denotes concatenation (or another agreed combination function).

The commit transaction does not execute \(m\). Instead, it records a commitment
to both the authorization secret and the intended action, thereby time-stamping
the future reveal without disclosing the action itself.

\subsubsection{Reveal Transaction}
After the commit transaction is verified to have been included in the blockchain, the user submits a second transaction revealing:
\[
(x, m)
\]

\subsubsection{Verification}
Nodes validate the authorization by checking:
\begin{itemize}
    \item $F(x)$ matches the committed value,
    \item $F(x \parallel m)$ matches the committed value,
    \item the action $m$ is well-formed and valid.
\end{itemize}

If all checks succeed, the action $m$ is executed. Each commitment is single-use:
after a successful reveal, the corresponding commitment is marked as spent and
cannot be used to authorize any subsequent state transition. This prevents
replay of the same preimage after it has become public.

This two-step process constitutes what we refer to as a \emph{commit--reveal (CR) signature}. Unlike conventional signatures, the CR signature is not a standalone object, but rather a pair of blockchain records whose consistency establishes authorization, with the finalized timestamp of the transaction interpreted as the time at which the commit transaction was accepted onto the blockchain.

\subsection{Security Intuition}

The security of the scheme follows directly from the preimage resistance of the function $F$. An adversary observing the commit transaction learns only the hash values $F(x)$ and $F(x \parallel m)$, and cannot derive $x$ without solving a preimage problem.

Even under quantum adversaries, the best known attacks against hash functions provide at most a quadratic speedup (e.g., Grover's algorithm), preserving strong security margins when appropriate hash lengths are used.

Thus, the scheme achieves post-quantum security without relying on lattice-based or multivariate assumptions.

\subsection{State Transition and Key Evolution}

A key consideration is that revealing $x$ exposes the corresponding authorization key. Therefore, after a reveal transaction, any remaining assets associated with $F(x)$ must be transferred to a fresh commitment.

To address this, the reveal transaction includes a new commitment:
\[
y' = F(x')
\]
where $x'$ is a newly generated secret preimage.

Upon execution, any remaining assets are reassigned to $y'$, ensuring that each preimage is used only once. This mechanism is analogous to key rotation and ensures forward security of remaining assets.

\subsection{Compact Representation}

The commit transaction can be optimized by combining hash outputs using a compression function. For example:
\[
C = F(x) \oplus F(x \parallel m)
\]
where $\oplus$ denotes bitwise XOR.

This reduces on-chain data size while preserving security, provided the resulting composite function remains collision-resistant.

\subsection{Discussion}

The proposed construction replaces a single large signature with two compact transactions, each containing fixed-size hash outputs. In contrast to post-quantum signature schemes that require kilobytes of data per signature, the CR scheme maintains a constant and minimal data footprint.

The key insight is that authorization can be expressed as a temporal relationship between two transactions, rather than as a static signature object. This shift allows post-quantum security to be achieved without incurring the large data overhead associated with conventional post-quantum signatures. Additional extensions, including multi-party authorization, atomic transactions, and cross-chain validation, can be constructed on top of this basic mechanism.

\section{Conclusion}

Post-quantum migration in blockchain systems is often framed as a problem of replacing vulnerable cryptographic primitives with quantum-resistant alternatives. This work shows that such a framing is fundamentally incomplete. In globally replicated systems, the dominant constraint is not cryptographic feasibility, but system-wide cost. Large post-quantum signature schemes introduce substantial increases in transaction size that are multiplied across all participating nodes, resulting in orders-of-magnitude growth in storage, bandwidth, and validation requirements. Even under conservative assumptions, this implies infrastructure expansion on the scale of tens of billions of dollars and necessitates multiple generations of hardware improvement to be deployed within compressed timeframes.

Crucially, these costs are not borne by a centralized operator or directly compensated by the protocol. Full nodes are typically not financially rewarded for storing and verifying the blockchain, relaying transactions, or supporting new nodes joining the network. As a result, increased resource requirements translate directly into higher uncompensated costs for participants. This creates a structural economic pressure toward reduced participation and consolidation among well-capitalized entities, with the potential to fundamentally alter the decentralized nature of blockchain systems and undermine their core value proposition.

The implication is clear: naïvely substituting large post-quantum signature schemes into existing transaction models is unlikely to be viable at scale. Instead, post-quantum security must be achieved within the constraints imposed by globally replicated infrastructure. To this end, we proposed a hash-based commit--reveal construction that eliminates large signature payloads and replaces them with fixed-size hash outputs. By shifting complexity from signature size to transaction structure, this approach bounds system-wide cost growth and remains compatible with incremental hardware evolution.

More broadly, these results suggest that post-quantum blockchain design must prioritize minimizing on-chain data expansion as a first-order objective. Approaches that ignore this constraint risk not only delayed adoption, but a transition toward systems that are operationally centralized in practice, regardless of their underlying protocol design. Future work should focus on formal security analysis, integration into existing blockchain environments, and empirical evaluation of deployment trade-offs, but the central conclusion remains: without careful attention to system-wide cost, post-quantum migration risks solving the cryptographic problem while breaking the system it is intended to secure.

\bibliographystyle{amsplain}
\bibliography{references}

\end{document}